\documentclass[preprint,superscriptaddress,showpacs]{revtex4}
\usepackage{dcolumn}
\usepackage{bm}
\usepackage{graphicx}
\usepackage{color}
\DeclareGraphicsExtensions{. jpg,. pdf, . mps, . png, . eps, . ps, . EPS}
\begin{document}
\def\be{\begin{equation}}
\def\ee{\end{equation}}

\def\bc{\begin{center}} 
\def\ec{\end{center}}
\def\bea{\begin{eqnarray}}
\def\eea{\end{eqnarray}}
\newcommand{\avg}[1]{\langle{#1}\rangle}
\newcommand{\Avg}[1]{\left\langle{#1}\right\rangle}
 \newcommand{\CHR}[1]{{\tt [#1 - CHR]}}

\title{A comparison between the quasi-species evolution and stochastic quantization of fields}
\author{ Ginestra Bianconi} 

\affiliation{Department of Physics, Northeastern University, Boston, 
Massachusetts 02115 USA}
\author{Christoph Rahmede}
\affiliation{Department of Physics, Technische Universit\"at Dortmund, D-44221 Dortmund, Germany}

\begin{abstract}
The quasi-species equation  describes the evolution of the probability that a  random individual in a population carries a given genome.
Here we map  the quasi-species equation for individuals of a self-reproducing population  to an ensemble of scalar field elementary units  undergoing a creation and annihilation process. 
In this mapping, the individuals of the population are mapped to field units and their genome to the field value.
The  selective pressure is mapped  to an inverse temperature $\beta$ of the system  regulating  the evolutionary dynamics of the fields.
We show that the quasi-species equation if applied to an ensemble of field units gives in the small $\beta$ limit can be put in relation with  existing stochastic quantization approaches.
The ensemble of field units described by the quasi-species equation relaxes to the fundamental state, describing an intrinsically dissipative dynamics.
For a quadratic dispersion relation the mean energy $\avg{U}$ of the system changes as a function of the inverse temperature $\beta$. For small values of $\beta$ the average energy  $\avg{U}$ takes a relativistic form, for large values of $\beta$, the average energy $\avg{U}$ takes a classical form.
\end{abstract}

\maketitle

\section{Introduction}
Increasing evidence shows that  the  Darwinian theory of biological  evolution \cite{Fisher,Nowak}  described by the quasi-species equation \cite{Eigen}  shares surprising similarities with quantum mechanics at the mathematical level \cite{Baake,Peliti,Leibler,GinChr2,Hanggi1, Hanggi2, Ebeling, GinChr1,Kingman,Kadanoff,BianconiO,Deem,Pastor_Satorras}.
 The quasi-species equation   describes the evolution of self-replicating macromolecules such as RNA or DNA or  asexual organisms.  In order to solve the quasi-species equation different methods from quantum mechanics have been used,  including quantum spin chains \cite{Baake}, path integrals \cite{Peliti,Leibler,GinChr2} and the Schr\"odinger equation in imaginary time \cite{Hanggi1,Hanggi2,Ebeling,GinChr1}. This intriguing relation extends also to the steady state of biological evolution  which is described by  a Bose-Einstein distribution in a number of evolutionary models  of  sexual and asexual populations \cite{Kingman,Kadanoff,BianconiO}.
Creation-annihilation operators and Fock-space formulations  have been used by several authors for modelling the stochastic dynamics of biological evolution \cite{Deem,Pastor_Satorras}.
Finally many-body theory approaches have been applied to the description of stochastic gene expression \cite{Wolynes}.\\
The relation between quantum mechanics and stochastic dynamics has been deeply explored over the  years \cite{Risken}. In particular the  existing theory of stochastic processes for diffusing particles is described by Langevin and Fokker-Planck equations. The latter  are known to be related to the solution of the Schr\"odinger equation in imaginary time. 
Thus the stochastic diffusion dynamics is used in the existing stochastic quantization approaches to define stochastic processes whose probability density converges to path integrals of some specific quantum system \cite{SQ,Mitter,sqbook}.
 \\
  Biological evolution is essentially a stochastic process determined by the reproductive rate (depending on the Fisher fitness of the individual and on the {\em selective pressure})  and by the mutation rate.
The resulting  stochastic  birth-death processes are described by the  quasi-species equation which is the mean-field equation for this stochastic dynamics.  The quasi-species equation for biological evolution is the simplest example for showing that many biological systems have an error threshold. If, in presence of a single-peak landscape, the mutation rate is higher than a threshold value the biological system is not able to climb to the fitness peak of the energy landscape. 
In addition to the mean-field description of biological evolution given by the quasi-species equation many interesting effects are due to the stochastic dynamics \cite{Mustonen},  fluctuating environment \cite{Leibler}, unbounded fitness landscapes \cite{Hallatschek}  or spatial structure and time delay effects \cite{MaynardSmith}.
\\
Already in 1975 J. B. Anderson \cite{Anderson} proposed to use a birth-death process. i.e. a Monte Carlo with cloning, to  reproduce the ground state of a given quantum system and the methods is currently used in the literature of optimization \cite{Martin1, Martin2}.
The starting point for our investigation is  that the quasi-species equation  defines  stochastic processes that in the small selection limit can be related to stochastic quantization \'a la Anderson \cite{Anderson}. These classes of stochastic processes involve creation-annihilation processes and a dissipative dynamics captured by the quasi-species equation with different   functions playing the role of the Fisher fitness and the  mutation rate.  
\\
In a previous paper \cite{GinChr1} we have investigated the mathematical structure of the quasi-species equation by applying it to an ensemble of particles undergoing a creation-annihilation process under the drive of an harmonic potential playing the role of the Fisher fitness and a Gaussian noise playing the role of the mutation rate. Moreover the  inverse temperature $\beta$ plays the role of the selective pressure.
This process can be viewed as an equivalent of the Ornstein-Uhlenbeck process in which the particles instead of diffusing are undergoing a creation-annihilation process.
In \cite{GinChr1} we found that the probability distribution of the process can be decomposed into eigenfunctions corresponding to a a discrete spectrum of eigenvalues. This spectrum is equivalent to  the spectrum of the quantum mechanical harmonic oscillator in the limit of small inverse temperature $\beta$.
\\
In this paper we extend  these results by investigating the evolution of an ensemble of elementary units, that we call  field units, associated with a scalar field $\rho(x)$. As individuals in a population carry a genome $\{\sigma_i\}_{i=1,\ldots,N}$ where $\sigma_i$ indicates the nucleotide at position $i$, the field units that we introduce in this paper, are  associated with a field $\rho(x)$. These field units might represent for example a strategy of an agent in an evolutionary game theory. The field units undergo a creation-annihilation process inspired by biological evolution.
A precise mapping can be done between the evolution of field units and biological evolution of asexual populations.
In this mapping each field unit corresponds to an individual of the population and   each value of the   field  corresponds  to the possible  genomes of the individual in the biological population. Moreover  the energy of the field unit corresponds to the Fisher fitness of biological evolution and a Gaussian noise corresponds to random mutations. Finally the inverse temperature $\beta$ corresponds to the selective pressure of biological evolution.
Solving the quasi-species equation for such a model implies that we need to find an expression for the probability $P(\{\rho(x)\})$ that a field unit carries a field $\rho(x)$. This quantity is a functional of $\rho(x)$ and therefore we need to extend   the quasi-species in order to describe the time evolution of  a functional.
For simplicity we solve explicitly the model for the energy of a  single field unit which is quadratic in the field strength, but the same framework might be extended to other form of the energy $U$.
We find that the distribution of the field units  can be decomposed in a discrete set of eigenvalues.
These eigenvalues are characterized by a set of discrete numbers $\{N_k\}$ extending in this way the stochastic quantization of particles found in \cite{GinChr1} to a stochastic quantization of fields.
Interestingly we find that in the limit of small inverse temperature $\beta $ the average energy  $\avg{U}$  takes a relativistic form while for high value of the inverse temperature the average energy $\avg{U}$ takes a classical form.
Finally the dynamics is an off-equilibrium dissipative dynamics that asymptotically in time  converges to the non-degenerated fundamental state $\{N_k=0\}$. This aspect of the evolutionary dynamics is especially interesting since it might allow application of this dynamics for devising new optimization algorithms.

\section{ The quasi-species equation of biological evolution } 
The genome of an asexual organism is formed by a single copy of each chromosome.
For a given population the genome is determined by the sequence $\{\sigma\}=(\sigma_1,\sigma_2,\ldots\sigma_i,\ldots \sigma_N)$ where $i=1,2,\ldots, N$ is the number of genetic loci, and each variable 
$\sigma_i$ can take  four possible values  $\sigma_i=1,2,3,4$ indicating the nucleotide at each genetic locus.
According to Darwinian evolution a biological  population evolves under the drive of selection and mutations.
We associate with a genome $\{\sigma\}$  the Fisher fitness $U(\{\sigma\})$ and a reproductive rate $W(\{\sigma\})=\exp[-\beta U(\{\sigma\})]$ where $\beta$ indicates the selective pressure.  
The selective pressure $\beta$  is such  that for $\beta=0$ all genotypes have the same reproductive rate $W(\{\sigma\})=1$, and when $\beta\gg1$ the different genotypes corresponding to different Fisher fitness $U(\{\sigma\})$ have very different reproductive rate. 
In order to simplify the model, we assume that the death rate is a random process independent of the genome of the individual. The quasi-species equation \cite{Nowak}  is the mean-field equation for the probability distribution $P(\{\sigma\},\tau)$ that at time $\tau$ an individual has genome $\{\sigma\}$ and is given by 
\begin{equation}
\frac{d P(\{\sigma\},\tau)}{d\tau}={{\mathbf M}_{\{\sigma\}|\{\sigma^{\prime}\}}\left[e^{-\beta  U(\{\sigma^{\prime}\})}P(\{\sigma^{\prime}\},\tau)\right]}-{Z_{\tau}}P(\{\sigma\},\tau)
 \label{Ev}
\end{equation}
where $Z_{\tau}$ is a normalization constant and can be expressed as  $Z_{\tau}=\Avg{e^{-\beta  U(\{\sigma\})}}_{\tau}$, i.e. as the  average over the probability distribution $P(\{\sigma\},\tau)$.
The operator ${\mathbf M}_{\{\sigma\}|\{\sigma^{\prime}\}}$ in Eq. $(\ref{Ev})$ describes the occurrence of mutations and acts on a generic function $h(\{\sigma^{\prime}\})$ as
\begin{equation}
{\mathbf M}_{\{\sigma\}|\{\sigma^{\prime}\}}h(\{\sigma^{\prime}\})=\sum_{\{\sigma^{\prime}\}}Q(\{\sigma\}
|\{\sigma^{\prime}\})h(\{\sigma^{\prime}\})\ .
\end{equation}
The matrix $Q(\{\sigma\}|\{\sigma^{\prime}\})$ represents the probability of mutations. 
If $\mu$ is the mutation rate we have 
\begin{equation}
Q(\{\sigma\}|\{\sigma^{\prime}\})=\prod_i\left[(1-\mu)\delta(\sigma_i, \sigma^{\prime}_i)+\frac{\mu}{4}\right]
\end{equation}
with $\delta(\sigma_i,\sigma_i^{\prime})$ indicating the Kronecker delta.

\section{ Evolution of an ensemble of particles } In \cite{GinChr1} we applied the quasi-species equation to  an ensemble of particles that are located in  one-dimensional space and that undergo a creation-annihilation process. 
The evolution of the particles can  be mapped to biological evolution. In this mapping, the fitness function corresponds to  the energy of a particle, and mutations correspond to  a stochastic noise. 
Finally  we assume that  the probability  to find a particle at a certain position  obeys the quasi-species equation.

If   $P(z,\tau)$ is the probability that a particle is at position $z$ at time $\tau$, the quasi-species equation for this ensemble of particles reads
\begin{equation}
\frac{d P(z,\tau)}{d\tau}={{\mathbf M}_{z,z^{\prime}}\left[e^{-\beta  U(z^{\prime})}P(z^{\prime},\tau)\right]}-{Z_{\tau}}P(z,\tau)
 \label{ev2}
\end{equation}
where the partition function $Z_{\tau}$ is given by
\begin{equation}
Z_{\tau}=\int dz^{\prime}\int dz \, Q(z,z^{\prime}) e^{-\beta  U (z^{\prime}))}P(z^{\prime})
\label{defZ}
\end{equation}
and the operator  ${\mathbf M}_{z,z^{\prime}}$, applied to a function $e^{-\beta  U(z)}f(z)$, acts as
\begin{equation}
{\mathbf M}_{z,z'}\left[e^{-\beta  U(z^{\prime})}f(z^{\prime})\right]=\int dz^{\prime} \, Q(z,z^{\prime})e^{-\beta  U(z^{\prime})}f(z^{\prime}) \nonumber
\label{M1}
\end{equation}
where $Q(z,z')$ describes the stochastic noise playing the role of { mutations} for  the evolution of the ensemble of particles.
The inverse temperature $\beta $,    with $\beta >0$, plays in this stochastic dynamics the same role as the selective pressure in the biological evolutionary dynamics.
For simplicity we assume that the position $z$ at time $t+dt$ is related to the position $z^{\prime}$ at time $t$ by $z=z^{\prime}+\eta$ where $\eta$ is a noise with Gaussian distribution. Therefore we take
\begin{equation}
Q(z,z^{\prime})=\sqrt{\frac{D}{2\pi\beta }}\int d\eta\, \delta(z^{\prime}-z-\eta)e^{-\frac{1}{2}\eta^2  D/\beta }
\label{Q2}
\end{equation}
where $\beta /D$ is the variance of the noise. 
In \cite{GinChr1} we studied this process and we found that the probability of a certain configuration of the ensemble of particles can be decomposed into eigenfunctions of the evolutionary operator corresponding to  a discrete spectrum of eigenvalues. For low values of $\beta $, the quasi-species equation is directly related to the Schr\"odinger equation in imaginary time.
In fact, introducing in Eq. $(\ref{M1})$ the quadratic expression $(\ref{Q2})$ and using the Fourier representation for the delta function we get
\begin{equation}
\hspace{-4mm}{\mathbf M}_{x,x^{\prime}}\left[e^{-\beta U(x^{\prime})}f(x^{\prime})\right]={\cal N}\int dx^{\prime} \int dk \, e^{-\beta[H(x,k)]+ik (x-x^{\prime})} f(x^{\prime})\nonumber
 \label{S}
\end{equation}
where ${\cal N}=\frac{1}{2\pi}\sqrt{\frac{1}{2\pi \beta D}}$ and the  Hamiltonian $H(x,k)$ of this system is identified as
$H(x,k)=\frac{D}{2 }k^2+U(x)$.
The action of the operator ${\mathbf M}_{x,x^{\prime}}$ on a function $f(x)$
for  $\beta \ll 1$ is given by
\begin{eqnarray}
&&\hspace*{-10mm}{\mathbf M}_{x,x^{\prime}}\left[e^{-\beta U(x^{\prime})}f(x^{\prime})\right]\simeq\nonumber\\
&&\simeq{\cal N}\int dx^{\prime} \int dk \left\{1-\beta H(x,k)\right\}e^{ik (x-x^{\prime})} f(x^{\prime})\ .\nonumber 
 \label{Sa}
\end{eqnarray}
Therefore for $\beta\ll1 $ the evolution Eq. (\ref{ev3}) can be written as 
\begin{eqnarray}
\hspace*{-10mm}\frac{d P(x,t)}{dt}&\simeq&-\frac{\cal N}{Z_t} \beta\left[-\frac{D}{2}\frac{\partial^2 P(x,t)}{\partial x^2 }+
U(x)P(x,t)\right] \nonumber \\
&&+\left(\frac{\cal N}{Z_t} -1\right)P(x,t)\ .
 \label{SA}
\end{eqnarray}
This equation is particularly interesting since it is linear in the Hamiltonian $H(x,k)$ and can be put in relation with the  Schr\"odinger equation in imaginary time. Therefore in this limit we are considering a quantum mechanical system evolving in imaginary time and we recover the classic model introduced by J. B. Anderson \cite{Anderson} in order to find the ground state of a quantum mechanical system.

 Nevertheless in the high-energy limit $\beta \gg1$ the quasi-species equation strongly deviates from the Schr\"odinger equation. The case of  an harmonic potential has been  explicitly solved in \cite{GinChr1} where it was  shown that the eigenvalues of the evolutionary operator in the $\beta \ll 1$ limit are given by the spectrum of the quantum harmonic oscillator.

\section{  Evolution of field units }
In this paper we apply  the quasi-species equation to a discrete ensemble of field units of a scalar field.
These field units generalize the concept of an individual of a population to a wider context. 
As the genome of an individual is defined by a sequence of variables $\{\sigma\}$ we assume that each field unit is associated to a  value of the scalar field $\rho(x)$ taking complex values and defined on a  continuous one-dimensional variable $x$.  As the individuals in a biological population undergo a birth-death process we assume that  the  field units undergo a creation-annihilation process and we suppose that they follow the quasi-species equation.
In the mapping between the population of asexual individuals and the ensemble of field units, the Fisher fitness is mapped to the  energy $U(\{\rho(x)\})$ associated to the value of the field, and the mutation probability is substituted with a Gaussian noise.
We might describe this dynamics in the framework of an evolutionary game context. We assume that the elementary units are individuals with a given strategy $\rho(x)$ and they play against a common player with fixed  strategy playing the role  of  a fixed environment. The probability that a strategy $\rho(x)$ reproduces depends on its fitness (payoff of the strategy in the given environment) $W(\{\rho(x)\})=e^{-\beta U(\{\rho(x)\})}$ and there is a random drift of the elementary units indicating that strategy might fade away if not successful. 
We indicate by the functional $P(\{\rho(x)\},\tau)$  the probability that at time $\tau$ a field unit  is associated with the field  $\rho(x)$. 
The quasi-species equation for the ensemble of field units is an  equation for the functional $P(\{\rho(x)\})$ given by
\begin{eqnarray}
\hspace*{-3mm}\frac{d P(\{\rho(x)\},\tau)}{d\tau}&=&{\mathbf M}_{\{\rho(x)\},\{\tilde{\rho}(x)\}}\left[e^{-\beta  U(\{\tilde{\rho}(x)\})}P(\{\tilde{\rho}(x)\},\tau)\right]-\nonumber \\
&&Z_{\tau} P(\{\rho(x)\},\tau)
 \label{ev3}
\end{eqnarray}
where
the partition function $Z_{\tau}$ is
\begin{eqnarray}
\hspace*{-12mm}Z_{\tau}&=&\int {\cal D}\rho(x)  {\cal D}\rho^{\star}(x)   {\cal D}\tilde{\rho}(x)   {\cal D}\tilde{\rho}^{\star}(x)\times\nonumber \\
&&\times Q(\{\rho(x)\},\{\tilde{\rho}(x)\})e^{-\beta  U (\{\tilde{\rho}(x)\})}P(\{\tilde{\rho}(x)\},\tau).
\label{Zt}
\end{eqnarray}
The operator  ${\mathbf M}_{\{\rho(x)\},\{\tilde{\rho}(x)\}}$ applied to a function $h(\tilde{\rho}(x))$ acts according to 
\begin{eqnarray}
\hspace*{-3mm}{\mathbf M}_{\{\rho(x)\},\{\tilde{\rho}(x)\}}\left[h(\tilde{\rho}(x))\right]&=& \int {\cal D }\tilde{\rho}(x) \int {\cal D }\tilde{\rho}^{\star}(x) \times\nonumber \\
&&\hspace*{-9mm}\times Q(\{\rho(x)\},\{\tilde{\rho}(x)\}) h(\{\tilde{\rho}(x)\})\ 
\label{M1}
\end{eqnarray}
where the kernel $Q(\{\rho(x)\},\{\tilde{\rho}(x)\})$ describes the stochastic noise playing the role of mutations.
The {\em inverse temperature} $\beta >0$ plays in this stochastic dynamics the same  role as the {\em selective pressure} in the biological evolutionary dynamics.
For simplicity we assume that  the function $Q(\{\rho(x)\},\{\tilde{\rho}(x)\})$ has the form of a noise with Gaussian distribution.
We assume that the field $\rho(x)$ is defined on a box of size $L$ with periodic boundary conditions. Therefore we take
\begin{equation}
Q(\{\rho(x)\},\{\tilde{\rho}(x)\})={\cal N}e^{-\frac{\cal C}{4\pi L}\int dx  [\tilde{\rho}^{\star}(x)-\rho^{\star}(x)][\tilde{\rho}(x)-\rho(x)]}
\label{Qx}
\end{equation}
where ${\cal N}$ is the normalization factor. The variance of the noise $Q(\{\rho(x)\},\{\tilde{\rho}(x)\})$,  given  by $1/{\cal C}=\beta  mc^2$, is proportional to the inverse temperature and to a typical energetic scale  of the system that we indicate by $mc^2$ \cite{nota}.
The energy $U(\{\rho(x)\})$ defines the probability that one field unit replicates and generates another field unit. The lower the energy $U(\{\rho(x)\})$ the higher is the probability that a field unit replicates.
By repeating the same argument used in the preceding paragraph, the  equation $(\ref{ev3})$ reduces to a Schr\"odiger equation for the field functionals only in the limit of small values of $\beta$. For large value of the selection pressure the dynamical equation we are considering strongly deviates from it.

In order to derive a specific solution of the stochastic dynamics of the ensemble of scalar field units described by Eq. $(\ref{ev3})$,   we take the energy quadratic in the field $\rho(x)$. In particular we take
\begin{equation}
U(\{\rho(x)\})=\frac{\hbar^2}{m}\frac{1}{4\pi L}\int dx {\cal L}(\rho(x)),
\label{Ux}
\end{equation}
with
\begin{equation}
 {\cal L}(\rho(x))=\left(\frac{\partial \rho^{\star}(x)}{\partial x}\right)\left(\frac{\partial \rho(x)}{\partial x}\right)+\frac{m^2c^2}{\hbar^2}\rho^{\star}(x)\rho(x).
 \label{cL}
\end{equation}
 The choice of this energy might be generalized to energy including higher powers of the field $\rho(x)$ or to consider adaptive fitness landscapes inspired  by models of pattern formation or models of  motion of active walker  driven by chemotaxis \cite{Schweitzer}. 
 Since $\rho(x)$ is defined on a box of size $L$, it  can be decomposed into a discrete series of Fourier components $\rho(k)$ associated with a $k=\frac{2\pi n}{L}$ that can be decomposed  according to   $\rho(k)=\rho_R(k)+i\rho_I(k)$ where $\rho_R(k),\rho_I(k)$ are the real and the imaginary parts of $\rho(k)$. 
The energy $U(\{\rho(x)\})=U(\{\rho(k)\})$ given by Eqs. $(\ref{Ux}),(\ref{cL})$ can be written as
\begin{equation}
U(\{\rho(k)\})=\frac{1}{2mc^2}\sum_k \epsilon^2(k)[\rho_{R}^2(k)+\rho_I^2(k)]
\label{Uk}
\end{equation}
where $\epsilon(k)$ is defined as 
\begin{equation}\label{epsilonofk}
\epsilon^2(k)=[(\hbar k c)^2+(mc^2)^2].
\end{equation}
We note that  we can express also the noise $Q(\{\rho(x)\},\{\tilde{\rho}(x)\})=Q(\{\rho(k)\},\{\tilde{\rho}(k)\})$, given by Eq. $(\ref{Qx})$, in terms of $\rho_R(k),\rho_I(k)$, obtaining
\begin{equation}
\hspace*{-5mm}Q(\{\rho(k)\},\{\tilde{\rho}(k)\})=\prod_k \prod_{\alpha=R,I}\sqrt{\frac{{\cal C}}{2\pi }} e^{-\frac{\cal C}{2}  [\rho_{\alpha}(k)-{\tilde{\rho}}_{\alpha}(k)]^2}.
\label{Qk}
\end{equation}
In order to solve the evolution of this ensemble of field units  given by  Eq. $(\ref{ev3})$, we study the eigenfunction problem
\begin{eqnarray}
{\mathbf M}_{\{\rho(x)\},\{\tilde{\rho}(x)\}}\left[e^{-\beta  U(\{\tilde{\rho}(x)\})}\pi_{\vec{n}}(\tilde{\rho}(x))\right]= \lambda_{\vec{n}} \pi_{\vec{n}}(\{\rho(x)\})
\label{eig}
\end{eqnarray}
where $\vec{n}=\{n_R(k_1),n_I(k_1),\ldots, n_R(k_L),n_I(k_L)\}$ characterize the eigenfunctions which are  solutions of the eigenvalue problem in Eq. $(\ref{eig})$.
The eigenfunction $\pi_{\vec{n}}(\{\rho(k)\})$ of the eigenvalue problem Eq. $(\ref{eig})$, with the choice of $U(\{\rho(k)\})$ given by Eq. $(\ref{Uk})$ and $Q(\{\rho(k)\},\{\tilde{\rho}(k)\})$  given by Eq. $(\ref{Qk})$ factorizes on functions defined on the single variables $\rho_{R,I}(k)$, i.e.  
\begin{equation} 
\pi_{\vec{n}}(\{\rho(k)\})=\prod_k \prod_{\alpha=R,I}\pi_{n_{\alpha}(k)}(\rho_{\alpha}(k))\ .
\label{pin}
\end{equation}
The single functions $\pi_{n_{\alpha}(k)}(\rho_{\alpha}(k))$ with $\alpha=R,I$ are given by 
\begin{equation}
\pi_{n_{\alpha}(k)}({\rho_{\alpha}(k)})=V_{n_{\alpha}(k)}(\rho_{\alpha}(k))e^{-\frac{1}{2}\gamma(k)\rho_{\alpha}^2(k)}
\label{eig}
\end{equation}
with  $V_{n_{\alpha}(k)}(\rho_{\alpha}(k))$ indicating a polynomial of order $n_{\alpha}(k)$. The only  possible value for $\gamma(k)$ that is positive and allows for a normalizable eigenfunction 
is given by 
\begin{equation}
\gamma(k)=\frac{\epsilon(k)}{2mc^2}\left(\sqrt{\Delta(k)}-\beta \epsilon(k)\right)
\label{gammak}
\end{equation}
with
\begin{equation}
\Delta(k)={\beta }^2\epsilon^2(k) +4\ .
\label{Deltak}
\end{equation}
\begin{figure}
\begin{center}
\includegraphics[width=0.6\columnwidth]{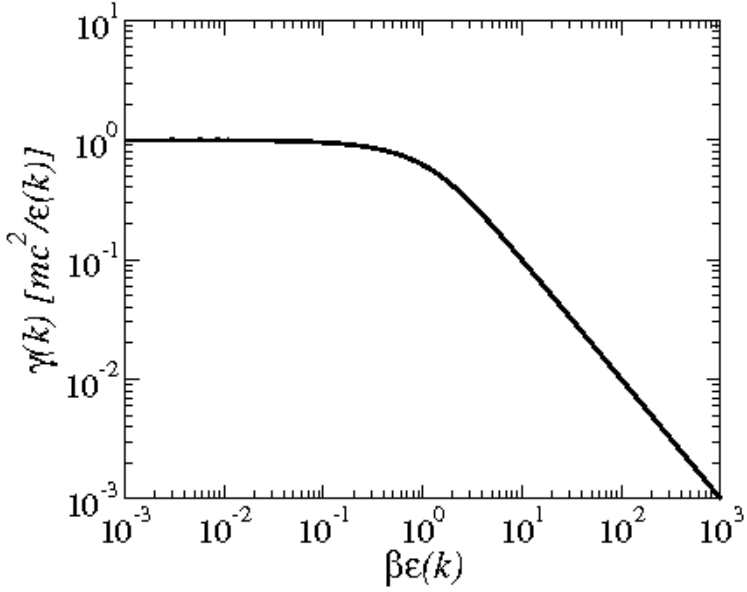}
\end{center}
\caption{The function $\gamma(k)$ defined in Eq. (\ref{gammaofk}) as a function of $\beta  \epsilon(k)$, where $\epsilon(k)$ has been defined in Eq. (\ref{epsilonofk}) and $\beta $ is the inverse temperature. }
\label{fig.gamma}
\end{figure}
In Figure $\ref{fig.gamma}$ we plot the function $\gamma(k)$ showing that as a function of $\beta \epsilon(k)$ the function is monotonically decreasing.
This implies that the distribution $\pi_{n_{\alpha}(k)}(\rho_{\alpha}(k))$ is  broader as $\beta$ increases and therefore the average energy $\avg{U(\{\rho(k)\})}$ will increase with $\beta$.
In the limit $\beta  \epsilon(k) \ll 1$ the function $\gamma(k)$ given by Eq. $(\ref{gammak})$ converges to its  maximum
\begin{equation}\label{gammaofk}
\gamma(k)=\frac{\epsilon(k)}{mc^2}+{\cal O}(\beta  \epsilon(k))\ .
\end{equation}
In the limit $\beta \epsilon(k) \gg1$ instead we find the asymptotic behavior
\begin{equation}
\gamma(k)=\frac{1}{\beta mc^2}\left[1-\frac{2}{\beta^2  \epsilon^2(k)}+{\cal{O}}\left(\frac{1}{{\beta}^4 \epsilon^4(k)}\right)\right].
\end{equation}
The eigenvalue $\lambda_{\vec{n}}$ of the eigenvalue problem Eq. $(\ref{eig})$ associated to the eigenfunction $\pi_{\vec{n}}(\{\rho(k)\})$ determined in Eq. $(\ref{pin})$ is given by   
\begin{equation}
\lambda_{\vec{n}(k)}=\prod_k{[1-\beta  mc^2\gamma (k)]}^{[n_R(k)+n_I(k)+1]}\nonumber \\
\end{equation}

\begin{figure}
\begin{center}
\includegraphics[width=0.6 \columnwidth]{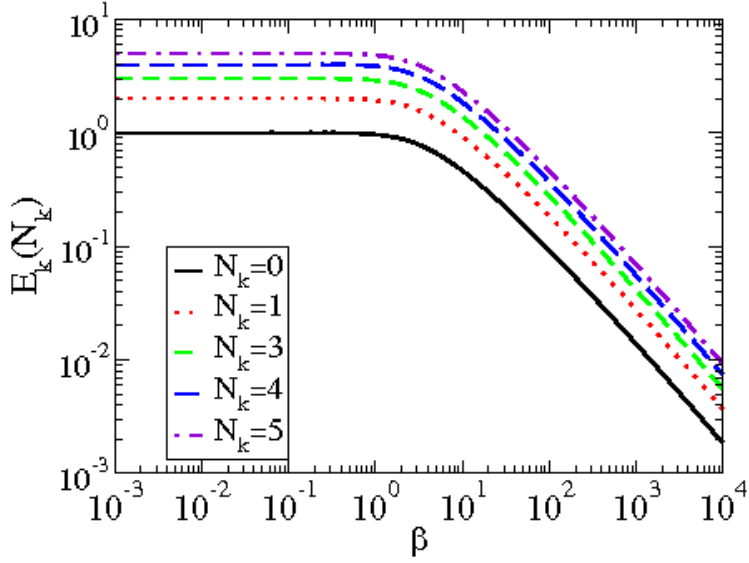}
\end{center}
\caption{The first five  levels $E_k(N_k)$ as a function of the inverse temperature $\beta $. The function is plotted for $\epsilon(k)=1$ which is defined in Eq. (\ref{epsilonofk}).}
\label{fig.1}
\end{figure}

\begin{figure}\begin{center}
\includegraphics[width=.6\columnwidth]{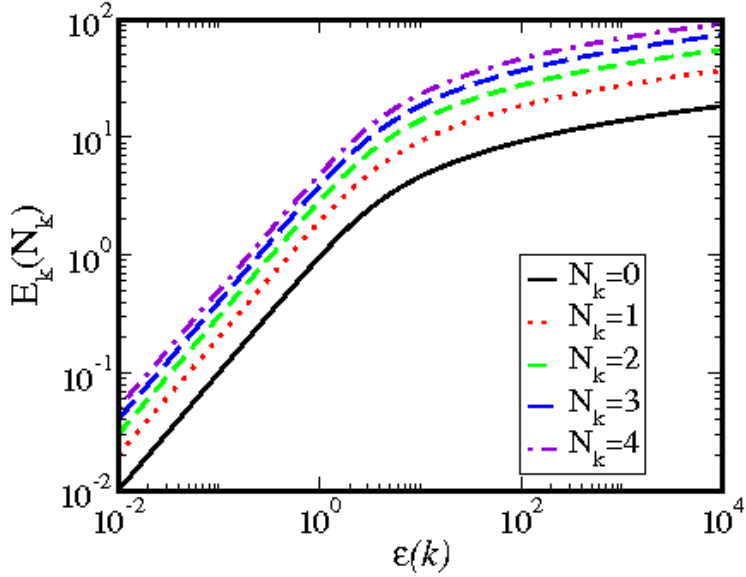}
\end{center}
\caption{The first five levels $E_k(N_k)$ as a function of  $\epsilon(k)$ defined in Eq. (\ref{epsilonofk}). The function is plotted for fixed inverse temperature $\beta =1$. }
\label{fig.2}
\end{figure}

Therefore the eigenvalues $\lambda_{\vec{n}}$ only depend on the variables $N_k=n_R(k)+n_I(k)$. 
We can write these eigenvalues as
\begin{equation}
\lambda_{\vec{n}(k)}=e^{-\beta  E(\{N_k\})}
\label{l}
\end{equation}
with 
\begin{eqnarray}
\hspace*{-3mm} E(\{N(k)\})&=&\sum_k  E_k(N_k)\nonumber \\
&=&-\frac{1}{\beta }\sum_k[N(k)+1]\ln[1-\beta  mc^2\gamma(k)].
\label{E}
\end{eqnarray}
Since  $N(k)=n_R(k)+n_I(k)$ each value $E_k(N_k)$ has a  degeneracy $g_k(N_k)= (N_k+1)$.
In Figure $\ref{fig.1}$ we plot $E_k(N_k)$ for $N_k=0,1,2,3,4$ versus   the inverse temperature $\beta $
while in Figure $\ref{fig.2}$ we plot the same functions versus the energy $\epsilon(k)$.
From these figures it is clear that the  levels  $E_k(N_k)$ of our model do not cross as a function of the {\it inverse temperature} $\beta$.
This is one of the major differences to traditional quasi-species models in which the level crossing is associated with the error-threshold phase transition.
While in biological evolution driven by a single-peak landscape,  the system of biologically replicating entities is not able to localize on the fitness peak if the mutation rate is too high, in our Gaussian model of evolution of field units, 
this phase transition does not occur and the maximal eigenvalue associated to  the fundamental state is an analytic function of $\beta$. This is dependent on the particular quadratic shape of the energy function we have assumed, and  other choices of the energy functional $U(\{\rho(x)\})$ might lead to a phase transition.
In the limit $\beta\epsilon(k) \ll 1$ the function $E(N_k)$ given by Eq. $(\ref{E})$ takes a relativistic form since using the definition of $\gamma(k)$ given by Eq. $(\ref{gammak})$ and $\Delta(k)$ given by Eq. $(\ref{Deltak})$ we get
\begin{equation}
E_k(N_k)=\sqrt{(\hbar kc)^2+(mc^2)^2}[N(k)+1)]
\end{equation}
with $N(k)=n_R(k)+n_I(k)$.
In the limit $\beta \epsilon(k) \gg1$ we obtain instead
\begin{equation}
E(\{N_k\})=\frac{2}{\beta }\sum_k (N_k+1)\ln\left[{\beta \epsilon(k)}/\sqrt{2}\right].
\end{equation}

Let us complete the solution of the stochastic evolution described by Eq. $(\ref{ev3})$.
First we  decompose the function $P(\{\tilde{\rho}(x)\},\tau)$ in the basis $\pi_{\vec{n}}(\{\rho(k)\})$ given by Eq. $(\ref{pin})$ and we get 
\begin{equation}
P(\{\rho(k)\},\tau)=\sum_{\vec{n}(k)} c_{\vec{n}(k)}(\tau) \pi_{\vec{n}}(\{\rho(k)\}).
\end{equation}
Since the dynamical Eq. $(\ref{ev3})$ can be easily  linearized as shown in \cite{Nowak}, the solution of Eq. $(\ref{ev3})$ is immediate. Using the definition of the eigenvalue $\lambda_{\vec{n}}=\exp[-\beta  E(\{N_k\})]$  given by Eq. $(\ref{eig})$ we get
\begin{equation}
c_{\vec{n}(k)}(\tau)=\exp\left[ e^{-\beta  E(\{N_k\})}\tau-{\cal Z}_{\tau}\right]c_{\vec{n}(k)}(0).
\label{dyn}
\end{equation}

The function ${\cal Z}_{\tau}$ appearing in Eq. $(\ref{dyn})$ and the function $Z_{\tau}$ defined in Eq. $(\ref{Zt})$ are given by 
\begin{eqnarray}
{\cal Z}_{\tau}&=&\int_0^{\tau} d\tau^{\prime}{Z_{\tau^{\prime}}} \label{def}\\
Z_{\tau}&=&\sum_{\vec{n}(k)}  e^{-\beta  E(\{n(k)\})} c_{\vec{n}(k)}(t)=\Avg{e^{-2\beta  E(\{n(k)\})}}\ \nonumber
\end{eqnarray}
where $\Avg{\dots}$ is the average over the ensemble of scalar field units.
The evolutionary process described by Eq. $(\ref{ev3})$ is an off-equilibrium and dissipative.
In fact using Eq. $(\ref{def})$ and the dynamical solution for $c_{\vec{n}(k)}$ given by Eq. $(\ref{dyn})$ we get 
\begin{equation}
\frac{dZ_{\tau}}{d\tau}=\left[\Avg{e^{-2\beta  E(\{n(k)\})}}-\Avg{e^{-\beta  E(\{n(k)\})}}^2\right]\ .
\end{equation}
Therefore,  asymptotically in time the ensemble of scalar fields concentrates on the maximal eigenvalue of the evolutionary dynamics, similarly to what happens in biological evolution under the fundamental theorem of natural selection \cite{Fisher}.
 In fact the ratio between the time dependent coefficients $c_{\vec{n}_1}(\tau),c_{\vec{n}_2}(\tau)$ follows the relation 
\begin{equation}
\hspace*{-5mm}\frac{c_{\vec{n}_1}(\tau)}{c_{\vec{n}_2}(\tau)}=\frac{c_{\vec{n}_1}(0)}{c_{\vec{n}_2}(0)}\exp\left\{\left[e^{-\beta  E(\{n_1(k)\})}-e^{-\beta  E(\{n_2(k)\})}\right]\tau\right\}\ .
\end{equation}
From Eq. $(\ref{dyn})$ it is easy to show that  the eigenfunction corresponding to the largest eigenvalue $\lambda_{\vec{n}(k)}$ and lower   value of $E(\{N_k\})$ is dominating at large time scales.
Interestingly,   if the probability distribution $P(\{\rho(k)\},0)$ has a non-zero contribution $c_{\vec{n}=\vec{0}}(0)>0$ of  the non-degenerate fundamental state $\{N_k\}=(0,0\ldots, 0)$,
the probability distribution asymptotically in time converges to the fundamental eigenfunction 
\begin{equation}
P(\{\rho(k)\},\tau)\to \prod_k \prod_{\alpha=R,I}\pi_{0}(\rho_{\alpha}(k))
\label{eigf}
\end{equation}
corresponding to the  value 
\begin{eqnarray}
E(\{N_k=0\})&=&-\frac{1}{\beta }\sum_k\ln[1-\beta  mc^2 \gamma(k)]\ .
\end{eqnarray}
If we impose a  cutoff  $K$ to the momentum $k$ (i.e. $k<K$ ) we get that in  the limit $\beta \epsilon(K) \ll1$, $E(\{N_k=0\})$ takes the form
\begin{eqnarray}
E(\{N_k=0\})&=&\sum_k\sqrt{(\hbar {k}c )^2+(mc^2)^2}\ .
\end{eqnarray}
In the opposite limit in which  $\beta  mc^2 \gg1$,  $E(\{N_k=0\})$ takes the form
\begin{equation}
E(\{N_k=0\})=\frac{2}{\beta } \sum_k \ln\left({\beta \epsilon(k)}/\sqrt{2}\right).
\end{equation}
The fundamental state is characterized by the eigenfunction in Eq. $(\ref{eigf})$ with $\pi_{0}(\rho_{\alpha}(k))$ given by a Gaussian as described by Eq. $(\ref{eig})$. Therefore, using the definition of $\gamma(k)$ given by Eq. $(\ref{gammak})$ we find
\begin{equation}
\avg{\rho_{\alpha}^2}=\frac{1}{\gamma(k)}
=\frac{2mc^2}{\epsilon(k)}\frac{1}{\sqrt{\Delta(k)}-\beta \epsilon(k)},
\label{rho2}
\end{equation}
with $\alpha=R,I$.
Finally, asymptotically in time, the average energy $\avg{U(\{\rho(k)\})}$  of the field units  is given by the average of $U(\{\rho(k)\})$ defined in Eq. $(\ref{Uk})$ on the fundamental eigenfunction given by Eq. $(\ref{eigf})$. Therefore using Eq. $(\ref{rho2})$ we find
\begin{equation}
\Avg{U(\{\rho(k)\})}=\sum_k\frac{2\epsilon(k)}{\sqrt{\Delta(k)}-\beta \epsilon(k)}\ .
\end{equation}
In the limit $\beta \epsilon(K)\ll 1$  we find that the average energy $ \Avg{U(\{\rho(k)\})}$ is linear in $\epsilon(k)$,
\begin{equation}
\Avg{U(\{\rho(k)\})}=\sum_k \sqrt{(\hbar kc)^2+(mc^2)^2}.
\end{equation}
Therefore in this limit,  we find that, asymptotically in time, the average energy of the field units $\Avg{U(\{\rho(k)\})}$ is linear in 
$\epsilon(k)$ acquiring in this way an apparently relativistic expression.
In the limit $\beta mc^2\gg 1$ instead, we find that $\Avg{U(\{\rho(k)\})}$ is quadratic in $\epsilon(k)$,
 \begin{equation}
 \Avg{U(\{\rho(k)\})}=\beta   \sum_k \left[(\hbar kc)^2+(mc^2)^2\right].
\end{equation}
This shows that  the  stochastic quantization we have explored allows both for relativistic and non-relativistic expression of the average energy $ \Avg{U(\{\rho(k)\})}$.

\section{ Conclusions } 
Erwin Schr\"odinger  \cite{Schroedinger}   proposed that the complexity of living systems is related to   quantum mechanics. Since then this idea has fascinated both physicists and biologists \cite{Lloyd,Penrose}.  From the publication of {\em What is life?} however more than sixty years have passed and we still lack a solid scientific basis for  Schr\"odinger's proposal. However the question  {\it Is life physics?} \cite{Goldenfeld} is attracting the interest of an increasing number of scientists.
The connection between the stochastic dynamics, which lies at the heart of biological evolution,  and  quantum-like behavior opens the way to apply the powerful techniques of the formalism of quantum mechanics within the theory of evolution. Recently the literature on this topic is acquiring  a  certain momentum and  path-integrals, Schr\"odinger equation in imaginary time, Fock formalism and Bose-Einstein condensation are techniques which have been applied to  study the quasi-species equation and stochastic effects in biological evolution \cite{Baake,Peliti,Leibler,GinChr2,Hanggi1, Hanggi2, Ebeling, GinChr1,Kingman,Kadanoff,BianconiO,Deem, Pastor_Satorras}.  These techniques do not solve all the complex aspects of biological evolution, i.e. biological evolution is not exclusively described by quantum formalism, but these techniques  have been demonstrated to be extremely useful for solving specific biological questions.
\\
Many  interesting works relate stochastic dynamics and quantum mechanics using  stochastic quantization  techniques \cite{SQ,Mitter,sqbook,Anderson}.
In this paper  we have investigated at the same time the stochastic nature of biological evolution and the relation of this stochastic dynamics with    stochastic quantization. 
The quasi-species equation  describes the evolution of the probability that a  random individual in a population carries a given genome.
Here we map  the quasi-species equation for individuals of a self-reproducing population  to an ensemble of scalar field elementary units  undergoing a creation and annihilation process. 
In this mapping the field units correspond to individuals and the field $\rho(x)$ to their genome.The selective pressure instead maps to an "inverse temperature" $\beta$ that parametrize the evolution of the field units.
We have solved the specific case in which the role of the Fisher fitness is played by a quadratic energy potential of the field and the mutations by Gaussian noise on the field values. 
The ensemble of field units described by the quasi-species equation relaxes to the fundamental state, describing an intrinsically dissipative dynamics.
We have presented interesting relations between this stochastic evolution and  quantum-like behavior  with the emergence of discrete eigenvalues for the particle probability distributions. For a quadratic dispersion relation the mean energy $\avg{U}$ of the system changes as a function of the inverse temperature $\beta$. For small values of $\beta$ the average energy  $\avg{U}$ takes a relativistic form, for large values of $\beta$, the average energy $\avg{U}$ takes a classical form.
\\
In future works we plan to study how this mean-field picture described by the quasi-species equation is modified by stochastic effects.

G.B.  acknowledges interesting discussions with  H. Goldberg, G. Jona Lasinio, M. Kardar, P. Nath, M. A. Nowak, and S. Redner.


\begin{thebibliography}{10}


\bibitem{Fisher}
{R. A. Fisher, } {\it The Genetical Theory of Natural Selection} (Clarendon, Oxford,1930) 
\bibitem{Nowak}
M. A. {Nowak, } {\it Evolutionary Dynamics} (Belknap Press, Cambridge, MA,2006) 
\bibitem{Eigen}
M. Eigen,     Die Naturwissenschaften {\bf 64}, {541} (1977).
\bibitem{Baake}
E. Baake, M.  Baake  and H. Wagner,   {Phys. Rev. Lett.} {\bf 78}, {559} (1997).
\bibitem{Leibler}
S. {Leibler  and E. Kussell, } {Proc. Natl. Aca. Sci USA} {\bf 107}, {13183} (2010).
\bibitem{Peliti}
L.{ Peliti, }  {EPL} { \bf 57}, {745} (2002).
\bibitem{GinChr2}
G. { Bianconi  and C.  Rahmede}, Chaos, Sol. Frac.  http://dx.doi.org/10.1016/j.chaos.2011.10.006  (2011).
\bibitem{Ebeling}
R. Feistel  and W. Ebeling,  {\it Evolution of Complex Systems} (Kluwer Academic Publishers, Berlin,1989) 
\bibitem{Hanggi1}
J.{ Dunkel, W. Ebeling, L. Shimansky-Geier  and P. H\"anggi}, { Phys. Rev. E} {\bf  67},{ 061118} (2003).
\bibitem{Hanggi2}
J. Dunkel, S. Hilbert, L. Shimansky-Geier and P. H\"anggi,   {Phys. Rev. E} {\bf  69}, { 056118} (2004).
\bibitem{GinChr1}
G. Bianconi   and C.  Rahmede, Phys. Rev. E {\bf 83}, 056104 (2011).
\bibitem{Kingman}
J. F. C. Kingman, 
 { J. Appl. Prob.} {\bf  15},{1} (1978).
\bibitem{Kadanoff}
S. N. Coppersmith, R. D.    Blanck  and L. P. Kadanoff, {Jour. Stat. Phys.} {\bf  97},{ 1999} (2004).
\bibitem{BianconiO}
G. { Bianconi  and O. Rotzschke},  {Phys. Rev. E}{\bf  82}, {036109} (2010).
  \bibitem{Deem}
J.-M.   Park and M. W. Deem,  Jour. of Stat. Phys. {\bf 125}, {975} (2006).
  \bibitem{Pastor_Satorras}
R. Pastor-Satorras  and R. Sol\'e, SantaFe working papers http://www.santafe.edu/media/workingpapers/01-05-024.pdf
\bibitem{Wolynes}
M. Sasai and P. G. Wolynes, Proceedings of the National Academy of Science, {\bf 100}, 2374 (2003).
\bibitem{Risken}
H. {Risken},  {\it The Fokker-Planck Equation} ({Springer-Verlag, Berlin,1996}) 
\bibitem{SQ}
{G.  Parisi   and Y.    Wu, }  {Scientia Sinica} {\bf  24}, {483} (1981). 
\bibitem{Mitter}
G. {Jona-Lasinio and P. K.   Mitter,  }  {Commun. Math. Phys.} {\bf 101}, {409} (1985). 
\bibitem{sqbook}
M. {Namiki},  {\it Stochastic quantization} ({Springer-Verlag,Berlin},1992) 

\bibitem{Mustonen}
V. Mustonen and M. L\"assig, {Proc. Natl. Aca. Sci.} {\bf 107}, {4248} (2010). 
\bibitem{Hallatschek}
O. Hallatschek,  Proc. Natl. Aca. Sci. {\bf 108}, 1783 (2011).

\bibitem{MaynardSmith}
J. Maynard Smith,  Proc. R. Soc. Lond. B {\bf 219}, 315 (1983). 
\bibitem{Anderson}
J. B. Anderson, J. Chem. Phys. {\bf 63} 1499 (1975).
\bibitem{Martin1}
N. Cerf and O. C. Martin, Phys. Rev. E {\bf 51}, 3679 (1995).
\bibitem{Martin2}
M. Rousset and G. Stoltz, J. Stat. Phys. {\bf 123}, 1251 (2006).
\bibitem{nota}
We observe that the constants $c$, $\hbar$ and $\beta $ in the paper are assumed to have the respective physical dimensions, but can be taken to be free parameters of the dynamics  of  the field units. 
On the contrary the time $\tau$ is an a-dimensional time, it can be told as a time variable   rescaled with the typical time scale of the system. 
\bibitem{Schweitzer}
F. Schweitzer and L. Schimansky-Geier, Physica A {\bf 206}, 359 (1994).
\bibitem{Schroedinger}
E. { Schr\"odinger },  {\it  What is life?: The physical aspect of the living cell} ({Cambridge University Press, Cambridge},1944)   
\bibitem{Lloyd}{ S. Lloyd},    { Nature Physics} {\bf 5}, {164} (2009).
\bibitem{Penrose}
{D.  Abbott, P.  Davies  and A. K.  Pati}, {\it  Quantum Aspects of Life } ({Imperial College Press London},2008) 
\bibitem{Goldenfeld}
{ N. Goldenfeld  and C.   Woese},Ann. Rev. Cond. Matt. Phys. {\bf 2}, 375 (2010).
\end{thebibliography}
\end{document}